\documentclass[12pt]{article}

\usepackage{amssymb}
\usepackage{latexsym,amsmath,amsthm,amsfonts}
\usepackage{cite} 
\usepackage{url}  
\usepackage{graphicx}
\usepackage{authblk}
\usepackage{color}
\usepackage{booktabs}

\newcommand{\real}{\mathbb{R}}

\newcommand{\Z}{\mathbf{Z}}

\newcommand{\bd}{\mathbf{d}}

\newcommand{\h}{\mathbf{h}}

\newcommand{\br}{\mathbf{r}}

\newcommand{\x}{\mathbf{x}}

\newcommand{\z}{\mathbf{z}}

\newcommand{\0}{\mathbf{0}}
\newcommand{\1}{\mathbf{1}}

\newcommand{\bDelta}{\boldsymbol{\Delta}}

\newcommand{\bOmega}{\boldsymbol{\Omega}}

\newcommand{\bSigma}{\boldsymbol{\Sigma}}

\newcommand{\bmu}{\boldsymbol{\mu}}

\newcommand{\brho}{\boldsymbol{\rho}}

\newcommand{\be}{\begin{eqnarray}}
\newcommand{\ee}{\end{eqnarray}}
\newcommand{\bes}{\begin{eqnarray*}}
\newcommand{\ees}{\end{eqnarray*}}
\newcommand{\bi}{\begin{itemize}}
\newcommand{\ei}{\end{itemize}}
\newcommand{\bea}{\begin{eqnarray}}
\newcommand{\eea}{\end{eqnarray}}

\newcommand{\bZ}{\mathbf{Z}}

\newcommand{\bGam}{\mathbf{\Gamma}}
\newcommand{\bp}{\mathbf{p}}
\newcommand{\bgam}{\mathbf{\gamma}}

\title{HT-eQTL: Integrative Expression Quantitative Trait Loci Analysis in a Large Number of Human Tissues}
\author[1]{Gen Li}
\author[2]{Dereje D.\ Jima}
\author[2,3]{Fred A.\ Wright}
\author[4]{Andrew B.\ Nobel}

\affil[1]{Department of Biostatistics, Mailman School of Public Health, Columbia University}
\affil[2]{Center for Human Health and the Environment and Bioinformatics Research Center, North Carolina State University}
\affil[3]{Department of Statistics and Biological Sciences, North Carolina State University}
\affil[4]{Department of Statistics and Operations Research and Department of Biostatistics, University of North Carolina at Chapel Hill}

\begin{document}

\date{}

\maketitle
\newpage

\begin{abstract} 
{\bf Background:} 
Expression quantitative trait loci (eQTL) analysis identifies genetic markers associated with the expression of a gene.
Most existing eQTL analyses and methods investigate association in a single, readily available tissue, such as blood.
Joint analysis of eQTL in multiple tissues has the potential to improve, and expand the scope of, single-tissue analyses.
Large-scale collaborative efforts such as the Genotype-Tissue Expression (GTEx) program are currently generating high quality data in a large number of tissues.
However, computational constraints limit genome-wide multi-tissue eQTL analysis.
{\bf Results:} 
We develop an integrative method under a hierarchical Bayesian framework for eQTL analysis in a large number of tissues.
The model fitting procedure is highly scalable, and the computing time is a polynomial function of the number of tissues.
Multi-tissue eQTLs are identified through a local false discovery rate approach, which rigorously controls the false discovery rate.
Using simulation and GTEx real data studies, we show that the proposed method has superior performance to existing methods in terms of computing time and the power of eQTL discovery.
{\bf Conclusions:}
We provide a scalable method for eQTL analysis in a large number of tissues.
The method enables the identification of eQTL with different configurations and facilitates the characterization of tissue specificity.
\end{abstract}

\section{Background}
Expression quantitative trait loci (eQTL) analyses identify
single nucleotide polymorphisms (SNPs) that are associated with the expression level of a gene.
A gene-SNP pair such that the expression of the gene is associated with the value of the SNP
is referred to as an eQTL.
One may view eQTL analyses as Genome-Wide Association Studies (GWAS) with  multiple molecular phenotypes.
Identification of eQTLs is a key step in investigating genetic regulatory pathways.
To date, numerous eQTLs have been discovered to be associated with human traits such as height and complex diseases such as Alzheimer's disease and diabetes \cite{cookson2009mapping,mackay2009genetics}.

With few exceptions, existing eQTL studies have focused on a single tissue; in human studies this tissue is
usually blood.  An important next step in exploring the genomic regulation of expression is to simultaneously study
eQTLs in multiple tissues.  Multi-tissue eQTL analysis can strengthen the conclusions of single tissue
analyses by borrowing strength across tissues, and can help provide insight into the genomic basis of
differences between tissues, as well as the genetic mechanisms of tissue-specific diseases.

Recently, the NIH Common Fund's Genotype-Tissue Expression (GTEx) project has undertaken a large-scale
effort to collect and and analyze eQTL data in multiple tissues on a growing set of human subjects, and there
has been a concomitant development of methods for the analysis of such data.
Flutre et al.\ \cite{flutre2013statistical} developed a Bayesian method to jointly analyze eQTLs in multiple tissues.
The method relies on a permutation test to evaluate significance levels.
Li et al.\ \cite{li2013empirical} developed an empirical Bayes approach called MT-eQTL (``MT" stands for multi-tissue).
The method uses an approximate expectation-maximization (EM) algorithm to fit the model, and controls the false discovery rate (FDR)
of eQTL detections by adaptively thresholding local false discovery rates derived from the fitted model.
Both \cite{flutre2013statistical} and \cite{li2013empirical} make use of a binary configuration vector, with
dimension equal to the number of available tissues, to describe, for each gene-SNP pair, the presence
or absence of association in each tissue.
Sul et al.\ \cite{sul2013effectively} proposed a Meta-Tissue method to combine linear mixed models with meta-analysis
to detect eQTL in multiple tissues. The method analyzes one gene-SNP pair at a time.
Initial analyses and conclusions of the GTEx project are described in
\cite{ardlie2015genotype}.  As part of this work, the multi-tissue eQTL methods of \cite{flutre2013statistical} and \cite{li2013empirical} were applied
to 9 human tissues with sample size greater than 80, focusing on local (cis)
pairs for which the SNP is within one mega-base (Mb) of the transcription start site  (TSS) of the gene.
The analysis found that most eQTLs are common across all tissues, though the effect size may
vary from tissue to tissue.  In addition, there are a small, but potentially interesting, set of eQTLs that
are present only in a subset of tissues, the most common cases being eQTLs that are present in only
one tissue, or present in all but one tissue.

As GTEx and related projects proceed, data are being collected from an increasing number of subjects,
and an increasing number of tissues.
In the current GTEx database (v6p), more than 20 tissues have a sample size greater than 150.
{\color{black}Existing eQTL analysis methods are limited in one way or another in performing simultaneous local eQTL analysis in a large number of tissues.
On the one hand, methods like \cite{flutre2013statistical} and \cite{li2013empirical} quickly become intractable as the total number of configurations grows exponentially in the number of tissues.
On the other hand, methods such as \cite{sul2013effectively} need to fit a model for each gene-SNP pair; the fitting time is impractical when the number of gene-SNP pairs and the number of tissues are large.}

In this paper, we develop an efficient computational tool, called HT-eQTL (``HT" stands for high-tissue), for joint eQTL analysis.
The method builds on the hierarchical Bayesian model developed in \cite{li2013empirical},
but the estimation procedure is significantly modified to address scaling issue associated
with a large number of tissues.  Rather than fitting a full model, HT-eQTL fits models for all pairs of tissues
in a parallel fashion, and then synthesizes the resulting pairwise models into a higher order model for all tissues.
To do this, we exploit the marginal compatibility of the hierarchical Bayesian model, which is not an obvious property and was proven in \cite{li2013empirical}.
An important innovation is that we employ a multi-Probit model and thresholding to deal with the exponentially growing configuration space.
{\color{black}The resulting model and fitting procedure can be efficiently applied to the simultaneous eQTL
analysis of 20-25 tissues.}
Empirical Bayesian methods for controlling false discovery rates in multiple hypothesis testing are developed.
We design testing procedures to detect different families of eQTL configurations.
We show that the eQTL detection power of HT-eQTL is similar to that of MT-eQTL,
and that both outperform the tissue-by-tissue approach, in a simulation study with a moderate number of tissues.
We also compare HT-eQTL with the Meta-Tissue method in the analysis of the GTEx v6p data.
This analysis shows that the methods have largely concordant results,
but that HT-eQTL gains additional power by borrowing strength across tissues.


\section{Methods}\label{sec:model}
In this section we describe the HT-eQTL method, beginning with a review
of the hierarchical Bayesian model and the MT-eQTL method in \cite{li2013empirical}, and then
describing our proposal on how to fit the Bayesian model in high-tissue settings.
Next, we describe a local false discovery rate based method for performing flexible eQTL inference.
Finally, we discuss a marginal test and a marginal transformation to check and improve the goodness
of fit of the model.

\subsection{Review: Bayesian Hierarchical Modal and MT-eQTL Procedure}
\label{subsec:MT}

Consider a study with $n$ subjects and $K$ tissues.  From each subject we have genotype data and measurements
of gene expression in a subset of tissues.  In many cases, covariate correction will be performed prior to analysis of
eQTLs.
For $k =1,\ldots, K$, let $n_k \leq n$ denote the number of subjects
contributing expression data from tissue $k$.
Let $\lambda = (i,j)$ be the index of a gene-SNP pair consisting of gene $i$ and SNP $j$, and
let $\Lambda$ be the set of all local (cis) gene-SNP pairs.
For $\lambda = (i,j) \in \Lambda$ and $k = 1,\ldots, K$, let
$r_\lambda(k)$ denote the sample correlation between the expression level of
gene $i$ and the (covariate corrected) minor allele frequency of SNP $j$ in tissue $k$, and
$\rho_\lambda(k)$ be the corresponding population correlation.  Define
$\br_{\lambda} = (r_\lambda(1), \ldots, r_\lambda(K))$ to be the vector of sample
correlations across tissues, and define the vector $\brho_{\lambda}$ of population
correlations in the same fashion.

Let $\Z_{\lambda} = \h(\br_{\lambda}) \cdot \bd^{1/2}$, where $\h(\cdot)$ is the entrywise Fisher transformation,
$\cdot$ is the Hadamard product, and $\bd$ is a $K$-vector whose $k$th component is the number of samples in tissue $k$ minus the
number of covariates removed from tissue $k$ minus $3$.
With proper preprocessing of the gene expression data, the vector $\Z_{\lambda}$ is approximately
multivariate normal \cite{mudholkar1976distribution} with mean $\bmu_\lambda = \h(\brho_\lambda) \cdot \bd^{1/2}$
and marginal variance one.  In particular, if $\rho_{\lambda}(k) = 0$ then the $k$th component of $\Z_{\lambda}$
has a standard normal distribution, and can therefore be used as a z-statistic for
testing $\rho_{\lambda}(k) = 0$ vs $\rho_{\lambda}(k) \neq 0$.  Thus we refer to $\Z_{\lambda}$ as a z-statistic vector.

The MT-eQTL model introduced in \cite{li2013empirical} is a Bayesian hierarchical model for the random vector
$\bZ_{\lambda}$.  The model can be expressed in the form of a mixture as
\be\label{MT-model}
\bZ_\lambda \sim \sum_{\bgam \in \{0,1\}^K} p(\bgam) \, \mathcal{N}_{K}\left(\ \bmu\cdot\bgam\ ,\ \bDelta+\bSigma\cdot\bgam\bgam'\ \right) .
\ee
The mixture in (\ref{MT-model}) is taken over the set $\{0,1\}^K$ of length $K$ binary vectors.  Each vector
$\bgam \in \{0,1\}^K$ represents a particular configuration of eQTLs across the $K$ available tissues:
$\bgam_k = 1$
if the gene-SNP pair indexed by $\lambda$ is an eQTL in tissue $k$, and $\bgam_k = 0$ otherwise.
{\color{black}We define Hamming class $m$ ($m=0,\cdots,K$) as the set of all binary $K$-vectors having $m$ ones, which correspond to all configurations in which there is an eQTL in $m$ tissues and no eQTL in $K-m$ tissues.  }
The first parameter $p(\cdot)$ is a probability mass function on $\{0,1\}^K$ with the interpretation that $p(\bgam)$ is the prior probability of the configuration $\bgam$.
{\color{black}The $K$-vector $\bmu$ characterizes the average true effect size of eQTLs in each tissue.}
The $K \times K$ correlation matrix $\bDelta$ captures the behavior of $\bZ_\lambda$ when no eQTLs are present
($\bgam = 0$): its diagonal entries are 1 due to variance stabilization, and its off-diagonal entries
reflect correlations arising from subject overlap between tissues.
{\color{black} The $K \times K$ covariance matrix $\bSigma$ captures the covariance structure arising from the underlying biology: its diagonal entries reflect tissue specific variation in true effect sizes, and its off-diagonal entries reflect covariance of true effect sizes between tissues due to tissue commonalities.}
Let $\theta=\{\bp, \bmu,\bDelta,\bSigma\}$ denote the set of unknown model parameters.

Under the model (\ref{MT-model}) the distribution of $\bZ_\lambda$ is a normal mixture with each component
corresponding to a specific eQTL configuration.
In particular, if $\bgam=\0$ ($\lambda$ is not an eQTL in any tissue) then
$\bZ_\lambda\sim\mathcal{N}_K(\0,\bDelta)$;
if $\bgam=\1$ ($\lambda$ is an eQTL in all tissues) then $\bZ_\lambda\sim\mathcal{N}_K(\bmu,\bDelta+\bSigma)$.
The true configuration vector for each gene-SNP pair $\lambda$ can be viewed as a latent variable.
The main goal of a statistical analysis is to obtain the posterior distribution of each latent variable, and to use it to make inferences about eQTL configurations in multiple tissues.

In order to make inference about configuration vectors, we first estimate the model parameters
$\theta=\{\bp, \bmu, \bDelta,\bSigma\}$.
In practice it is common to set the average effect size vector $\bmu$ to $\0$,
as minor alleles are equally likely to be associated with high or low expression, and we assume in what follows
that $\bmu = \0$.
The remaining parameters can be estimated within a maximum pseudo-likelihood framework, where
the pseudo-likelihood is defined as the product of the likelihoods of all considered gene-SNP pairs.
We note that factorizing the likelihood in this way ignores dependence between adjacent and nearby SNPs
arising from linkage disequilibrium.  However, our interest is not in the joint behavior of the vectors $\bZ_\lambda$
but in their {\em marginal} behavior, which is reflected in the mixture (\ref{MT-model}).
In particular, the parameters in Model \eqref{MT-model} determine, and are determined by,
the marginal distribution of the vectors $\bZ_\lambda$, and do not depend on joint distribution of the vectors $\bZ_\lambda$.

\begin{figure}[h!]
\centering
\includegraphics[width=2.5in]{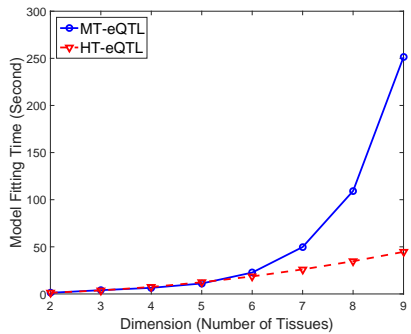}
\caption{The model fitting times of MT-eQTL and HT-eQTL for a sequence of nested models with dimensions 2 to 9 in the simulation study.The solid line with circles is for MT-eQTL, and the dashed line with triangles is for HT-eQTL.}\label{fig:2}
\end{figure}

A modified EM algorithm was devised in \cite{li2013empirical} to estimate the parameters from the pseudo-likelihood.
While the method scales linearly with sample size and the number of gene-SNP pairs,
its computational time increases exponentially with the number of tissues $K$ (see Figure \ref{fig:2}).
For genome-wide studies, it is infeasible to apply the method to data with more than a few tissues.
Moreover, the number of configurations grows exponentially with the number of tissues as well,
making inference about configurations difficult as well.
{\color{black}Below we introduce a scalable procedure, the HT-eQTL method, to address multi-tissue eQTL analysis in about 20 tissues.}

\subsection{The HT-eQTL Method}\label{HT}

The original MT-eQTL model has the desirable property of being {\em marginally compatible}.
Let the dimension of the MT-eQTL model be the number of available tissues.
Marginal compatibility means that: 1) the marginalization of a $K$-dimensional model to a subset of $L$ tissues has
the same general form as the $K$-dimensional model; and
2) the corresponding parameters for the $L$-dimensional model are obtained in the obvious way
by restricting the parameters of the $K$-dimensional model to the subset of $L$ tissues.

Because of marginal compatibility, it is straightforward to obtain a sub-model from a high dimensional model
without refitting the MT-eQTL parameters.
The HT-eQTL method, which is discussed below, estimates the high dimensional model
from the collection of its one- and two-dimensional sub-models.
Thus we address the computationally intractable problem of estimating a high dimensional model by considering
a manageable number of sub-problems that can be solved efficiently, and in parallel.

In the MT-eQTL model,
the covariance matrices $\bDelta$ and $\bSigma$ reflect interactions between pairs
of tissues, while the probability mass function $p(\cdot)$ captures higher order relationships between
tissues.  The HT-eQTL model is built from estimates of all one- and two-dimensional sub-models, which
can be computed in parallel.
In particular, we make use of a Multi-Probit model to approximate the $K$-th order probability mass function $p(\cdot)$ from the
probability mass functions of two-dimensional models.
In what follows we denote the estimated parameters of the two-dimensional model for tissue pair $(i,j)$ by
\[
\bp^{ij}=(p^{ij}_{00},p_{01}^{ij},p_{10}^{ij},p_{11}^{ij}),\
\bDelta^{ij}=\begin{pmatrix}
1 & \delta^{ij} \\
\delta^{ij} & 1                                                                                            \end{pmatrix},\
\bSigma^{ij}=\begin{pmatrix}
               \sigma_{11}^{ij} & \sigma_{12}^{ij} \\
               \sigma_{21}^{ij} & \sigma_{22}^{ij}
             \end{pmatrix}.
\]

{\noindent\bf Assemble $\bDelta$:}
{\color{black}For each tissue pair $(i,j)$ where $1 \leq i < j \leq K$, the corresponding off-diagonal value of $\bDelta$ is denoted by $\delta_{ij}$. An asymptotically consistent estimate of $\delta_{ij}$ is the off-diagonal
value of $\bDelta^{ij}$, which is the null covariance matrix for the two-dimensional model for tissue pair $(i,j)$. }
 Making this substitution for each $i < j$ and placing ones along the diagonal
yields a symmetric matrix $\hat{\bDelta}_0$.
If $\hat{\bDelta}_0$ is positive definite, it is directly used as an estimate $\hat{\bDelta}$ of $\bDelta$.
Otherwise (which never happened in our numerical studies), one could replace the non-positive eigenvalues of   $\hat{\bDelta}$ with 0.01 and rescale the matrix to have the diagonal entries equal to one.

{\noindent\bf Assemble $\bSigma$:}
To estimate the covariance matrix $\bSigma = \{ \sigma_{ij} \}$, we decompose it into the diagonal
values, which are tissue-specific variances, and the corresponding correlation matrix.
For each diagonal entry $\sigma_{kk}$ ($k=1,\cdots, K$), there are $K-1$ estimates, namely
$\sigma^{1k}_{22},\cdots,\sigma^{(k-1)k}_{22},\sigma^{k(k+1)}_{11},\cdots,\sigma^{kK}_{11}$.
In practice, the distribution of z-statistics is usually heavy-tailed, inflating the pairwise estimates of the variance.
As a remedy, we propose to use the minimum of the $K-1$ estimates as the estimate of $\sigma_{kk}$ to compensate the inflation effect. 
The induced correlation matrix from $\bSigma$ is estimated in the same way as $\bDelta$.
In particular, we begin with a matrix having ones along the diagonal, and off-diagonal entries
$\sigma_{12}^{ij} / \sqrt{\sigma_{11}^{ij} \sigma_{22}^{ij}}$.  After resetting any non-positive eigenvalue to 0.01, we rescale the matrix to have diagonal values equal to 1, and combine it with the variance estimates to obtain the estimate $\hat{\bSigma}$.

{\noindent\bf The Multi-Probit Model for $\bp$:}
Existing multi-tissue eQTL studies \cite{li2013empirical,ardlie2015genotype} support several broad conclusions
about eQTL configurations across tissues.  Researchers found that most gene-SNP pairs were not an eQTL in {\em any} tissue (Hamming class $0$) or were an eQTL in {\em all} tissues (Hamming class $K$).
With larger sample sizes and a larger number of tissues (thus providing increased power to detect cross-tissue sharing), we expect these two Hamming classes to predominate.

In general, the probability mass functions obtained from two-dimensional models will not determine a unique
probability mass function on the full $K$-dimensional model.  Here we make use of
a multi-Probit model through which we equate the values of the estimated probability mass function
with integrals of a multivariate normal probability density.
In particular, for each tissue pair $(i,j)$, we select thresholds $\tau^{ij}_1, \tau^{ij}_2 \in \real$
and a correlation $\omega^{ij} \in (0,1)$ so that if $(W_i, W_j)$ are bivariate normal with mean zero,
variance one, and correlation $\omega^{ij}$ then
\[
\Pr\left\{ \mathbb{I}(W_i \geq \tau^{ij}_1) = u \ \mbox{ and } \, \mathbb{I}(W_j \geq \tau^{ij}_2) = v \right\} = p^{ij}(u,v)
\]
for each $u, v \in \{0,1\}$.  Here $\mathbb{I}(A)$ is the indicator function of $A$, and $p^{ij}(\cdot)$ is the estimated
probability mass function for the pair $(i,j)$.

Beginning with a symmetric matrix having diagonal values 1 and off-diagonal values equal to $\omega^{ij}$,
we define a correlation matrix $\bOmega$ following the procedure used to define
$\hat{\bDelta}$.  Let $\phi_K(\cdot)$ be the probability density function of the corresponding $K$-variate normal
distribution $\mathcal{N}_K({\bf 0}, \bOmega)$.
{\color{black}For each tissue $j$, we define an aggregate threshold $\tau^j$ to be the minimum of
$\tau^{ij}_1$ ($i<j$) and $\tau^{ji}_2$ ($j<i$). Here we use the minimum because pairwise models may occasionally overestimate the null prior probability $p^{ij}(0,0)$.
Subsequently, for each configuration $\bgam \in \{0,1\}^K$,  we define the probability
\[
p(\bgam) = \int_{I_1}\cdots\int_{I_K} \phi_K(\x)d\x
\]
where $I_k$ is equal to $(-\infty, \tau^k]$ if $\gamma_k = 0$, and $(\tau^k, \infty)$, if $\gamma_k = 1$.
Consequently, we obtain the estimate of probability mass function $\bp$ for the $K$-dimensional model.}

{\noindent\bf Threshold $p(\cdot)$:}
In practice, many of the $2^K$ possible configurations will have estimated probabilities close to zero.
In order to further reduce the number of configurations, we set the threshold for the prior probabilities to be  $10^{-5}$, and truncate those values below the threshold to be zero.
The remaining probabilities are rescaled to have total mass one.
As a result, the total number of configurations with non-zero probabilities is dramatically reduced to a manageable level for subsequent inferences.

\subsection{Inferences}\label{sec:inference}

The first, and often primary, goal of eQTL analysis in multiple tissues is to detect which gene-SNP pairs are
an eQTL in some tissue.  Subsequent testing may seek to identify gene-SNP pairs that are an eQTL
in a specific tissues, and pairs that are an eQTL in some, but not all, tissues.
As the model \eqref{MT-model} is fit with large number of gene-SNP pairs, we ignore the estimation error
associated with the model parameters and treat the estimated values as fixed and true for the purposes
of subsequent inference.

The mixture model \eqref{MT-model} may be expressed in an equivalent, hierarchical form, in which
for each gene-SNP pair $\lambda$, there is a latent random vector
$\bGam_\lambda\in\bGam$ indicating whether or not that pair is an eQTL in each of the $K$ tissues.
The prior distribution of $\bGam_\lambda$ is characterized by the probabilistic mass function $p(\cdot)$.
In the hierarchical model, given that $\bGam_\lambda=\bgam$, the random z-statistic vector
$\bZ_\lambda$ has distribution
$
\mathcal{N}_K(\0,\bDelta+\bSigma\cdot\bgam\bgam')
$.
The posterior distribution of $\bGam_\lambda$ given the observed vector $\z_\lambda$ can be used to test eQTL
configurations for the gene-SNP pair $\lambda$.

Detection of eQTLs with specified configurations can be formulated as a multiple testing problem, and addressed
through the use of local false discovery rates derived from the posterior distribution of gene-SNP pairs.
Suppose that we are interested in identifying gene-SNP pairs with eQTL configurations in a set $S \subseteq \{0,1\}^K$.
This can be cast as a multiple testing problem
\bes
\mbox{H}_{0,\lambda}:\bGam_\lambda\in S^c \ \mbox{ versus }\ \mbox{H}_{1,\lambda}:\bGam_\lambda\in S
\ees
where $\lambda \in \Lambda$.
Rejecting the null hypothesis for a gene-SNP pair $\lambda$ indicates that $\lambda$ is likely to have an eQTL configuration
in $S$.  There are several families $S$ of particular interest, corresponding to different configurations of interest:
\bi

\item Testing for the presence of an eQTL in any tissue: \
$S=\{\bgam: \bgam \neq \0 \}$

\item Testing for presence of a tissue-specific eQTL, i.e., an eQTL in some, but not all, tissues: \
$S=\{\bgam: \bgam\neq \0, \bgam \neq\1 \}$

\item Testing for presence of an eQTL in tissue $k$ only: \  $S=\{\bgam: \gamma_k=1 \}$

\item Testing for presence of a common eQTL, i.e., an eQTL in all tissues: \ $S = \{\1\}$.

\ei

To carry out multiple testing under the hierarchical Bayesian model, we make use of the local false discovery rate (lfdr) for the set $S$, which is defined as the posterior probability that the configuration $\bGam$ lies in $S^c$ given the observed z-statistics vector $\z$.
The local false discovery rate was introduced by \cite{efron2001empirical} in the context of an empirical Bayes analysis of differential expression in microarrays. Other applications can be found in
\cite{newton2004detecting,efron2007size,efron2008microarrays}.
Formally, the lfdr for $S \subseteq \{0,1\}^K$ is defined by
\be\label{lfdr}
\eta_S(\z)
:=
\Pr(\bGam \in S^c \ | \ \bZ = \z) =
{ \sum_{\bgam \in S^c} p(\bgam) f_\gamma (\z) \over \sum_{\bgam \in \{0,1\}^K} \,
p(\bgam) f_\gamma (\z)},
\ee
where $f_\gamma(\z)$ is the pdf of $\mathcal{N}_K(\0,\bDelta+\bSigma\cdot\bgam\bgam')$.
Thus $\eta_S(\z_\lambda)$ is the probability of the null hypothesis given the z-statistic vector for the gene-SNP
pair $\lambda$.  Small values of the lfdr provide evidence for the alternative hypothesis $\mbox{H}_{1, \gamma}$.
In order to control the overall false discovery rate (FDR) for the multiple testing problem across all gene-SNP
pairs $\lambda \in \Lambda$ we employ an adaptive thresholding procedure for local false discovery rates
\cite{efron2001empirical, newton2004detecting, li2013empirical,sun2007oracle}.
For a given set of configurations $S$, and a given false discovery rate threshold $\alpha \in (0,1)$,
the procedure operates as follows.
\bi

\item Calculate the lfdr $\eta_S(\z_\lambda)$ for each $\lambda \in \Lambda$.

\item Sort the lfdrs from smallest to largest as $\eta_s(\lambda_{(1)}) \leq \cdots \eta_s(\lambda_{(N)})$.

\item Let $N$ be the largest integer such that
\[
{1 \over N} \sum_{i=1}^N \eta_s(\lambda_{(i)}) < \alpha.
\]

\item
Reject hypotheses $\mbox{H}_{0,\lambda_{(i)}}$ for $i = 1,\ldots, N$.

\ei
It is shown in \cite{li2013empirical} that the adaptive procedure controls the FDR at level $\alpha$ under
very mild conditions.
Consequently, we obtain a set of discoveries with FDR below the nominal level $\alpha$.

\section{Results}

In the first part of this section, we conduct a simulation study with 9 tissues.
{\color{black}We compare HT-eQTL with the MT-eQTL \cite{li2013empirical}, Meta-Tissue \cite{sul2013effectively} and tissue-by-tissue (TBT) \cite{heinzen2008tissue,dimas2009common,ding2010gene,fu2012unraveling} methods on different eQTL detection problems.
The Meta-Tissue approach leverages the fixed effects and random effects method to address effect size heterogeneity and detect eQTLs across multiple tissues.
The TBT approach
first evaluates the significance of gene-SNP association in each tissue separately, and then aggregates the information across tissues.
We also compare HT-eQTL and MT-eQTL in terms of the model fitting times and parameter estimation accuracy.}
{\color{black}Then we apply the two scalable methods, HT-eQTL and Meta-Tissue, to the GTEx v6p data with 20 tissues.}

\subsection{Simulation}
In the simulation study, we generate z-statistics directly from Model \eqref{MT-model} with $K=9$ tissues, using the MT-eQTL model parameters estimated from the GTEx pilot data \cite{li2013empirical}.
{\color{black}More specifically, for each gene-SNP pair, we first randomly generate a length-$K$ configuration vector according to the prior probability mass function $\bp$, and then simulate a $K$-vector of effect sizes  from the corresponding multivariate Gaussian distribution.
We simulate $10^5$ gene-SNP pairs in total. 
The true eQTL configurations for the simulated data are known and used to compare the efficacy of different methods.}

We first compare the computational costs of the MT-eQTL model fitting and the HT-eQTL model fitting (without parallelization).
We consider a sequence of nested  models with dimensions from 2 to 9.
The model fitting times {\color{black} on the simulated data }are shown in Figure \ref{fig:2}.
We demonstrate that the model fitting time for the MT-eQTL grows exponentially  in the number of tissues, while it grows much slower for the HT-eQTL.
Namely, the HT-eQTL scales better than the MT-eQTL.
This is because the HT-eQTL model fitting only involves the fitting of all the 2-tissue MT-eQTL models and a small overhead induced by assembling the pairwise parameters.
When the total number of gene-SNP pairs and the number of tissues are large, the advantage of HT-eQTL is significant.
{\color{black}Based on the timing results for MT-eQTL on the 9-tissue GTEx pilot data in \cite{li2013empirical}, we project its fitting time to be more than 30 CPU years on 20 tissues.
As we describe later, fitting the HT-eQTL model on the 20-tissue GTEx v6p data only takes less than 3 CPU hours.
We remark that the straightforward parallelization of the 2-tissue MT-eQTL model fittings will further reduce the computational cost for HT-eQTL.}

\begin{figure}[h!]
\centering
\includegraphics[width=2.3in]{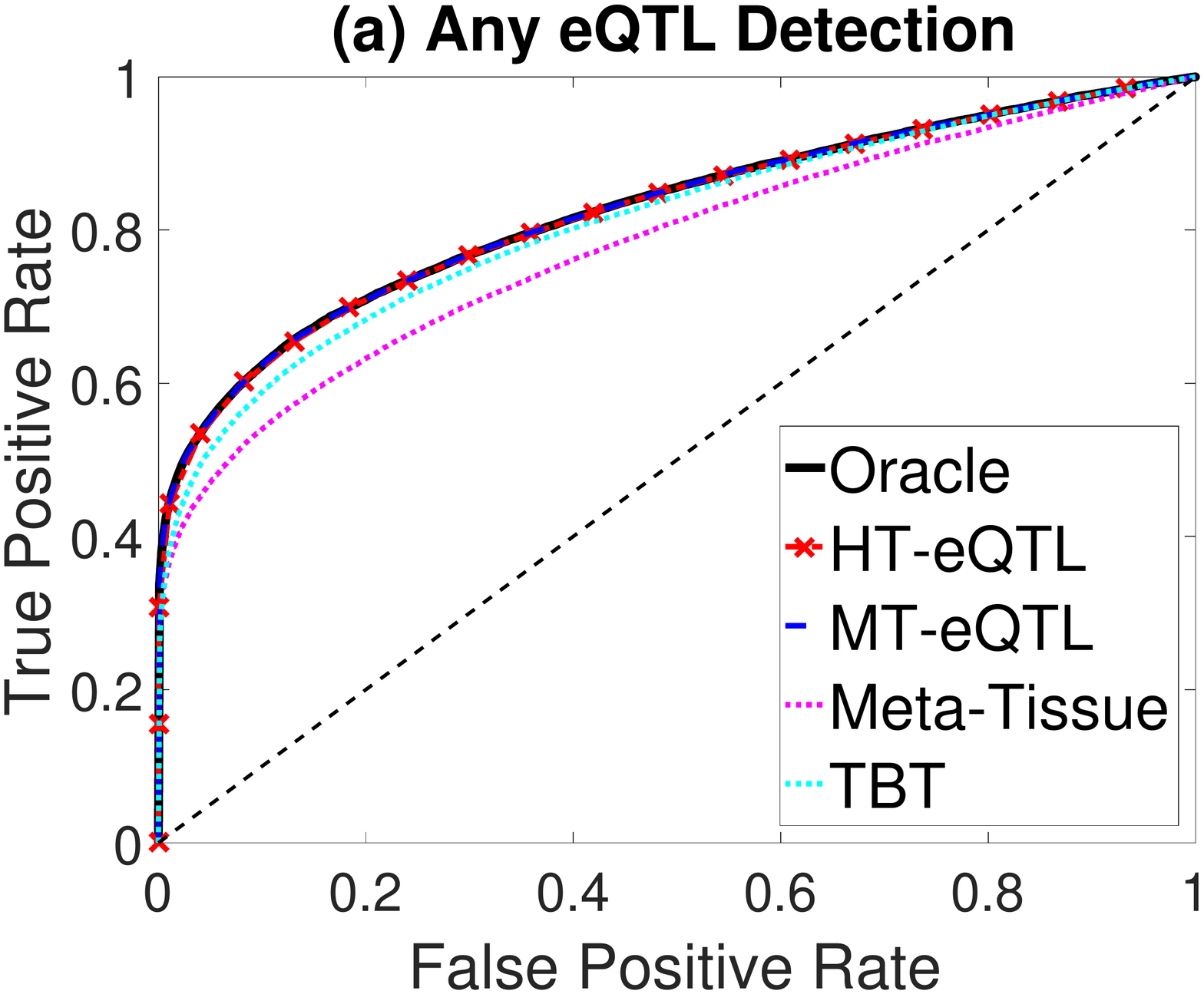}
\includegraphics[width=2.3in]{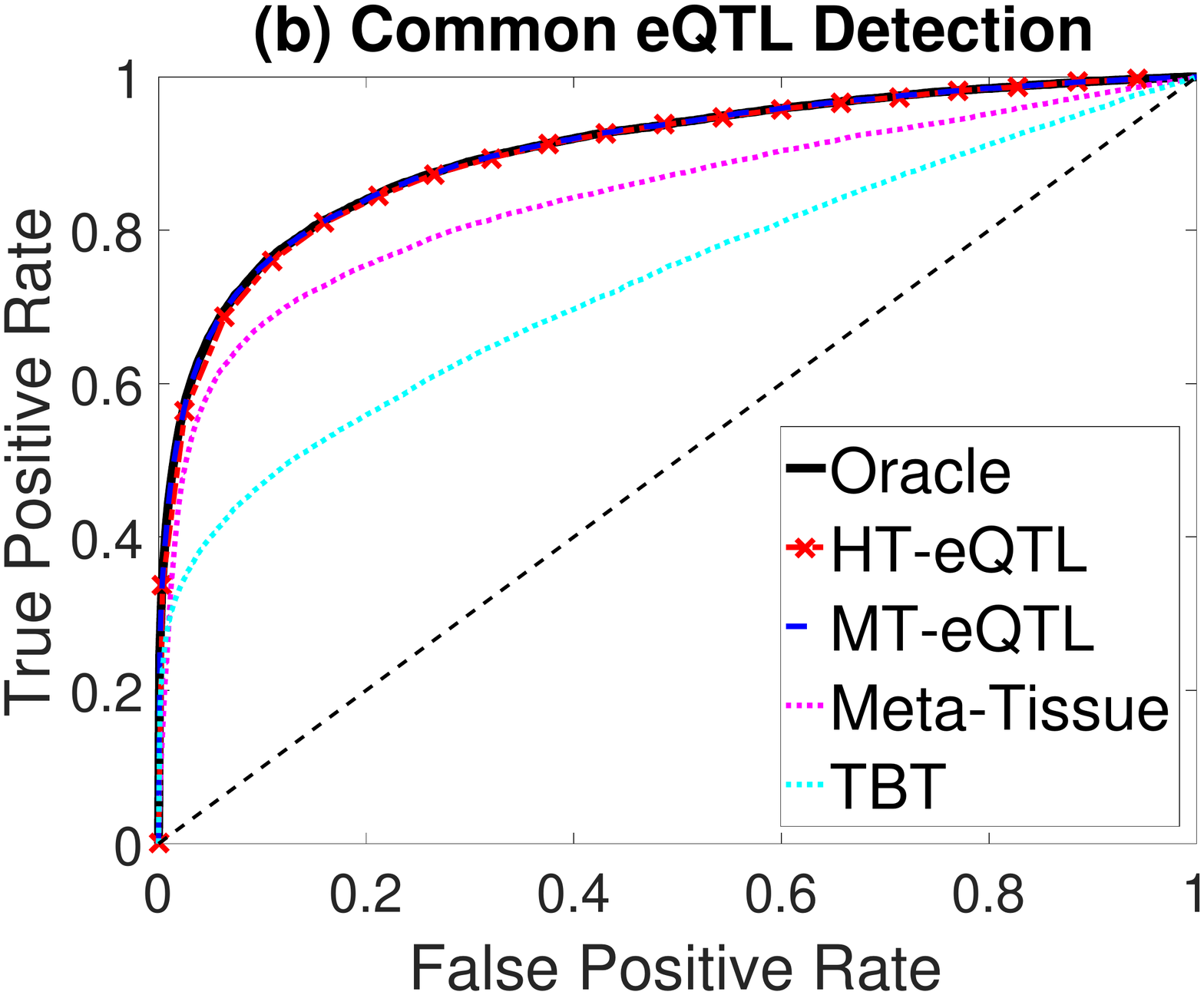}
\includegraphics[width=2.3in]{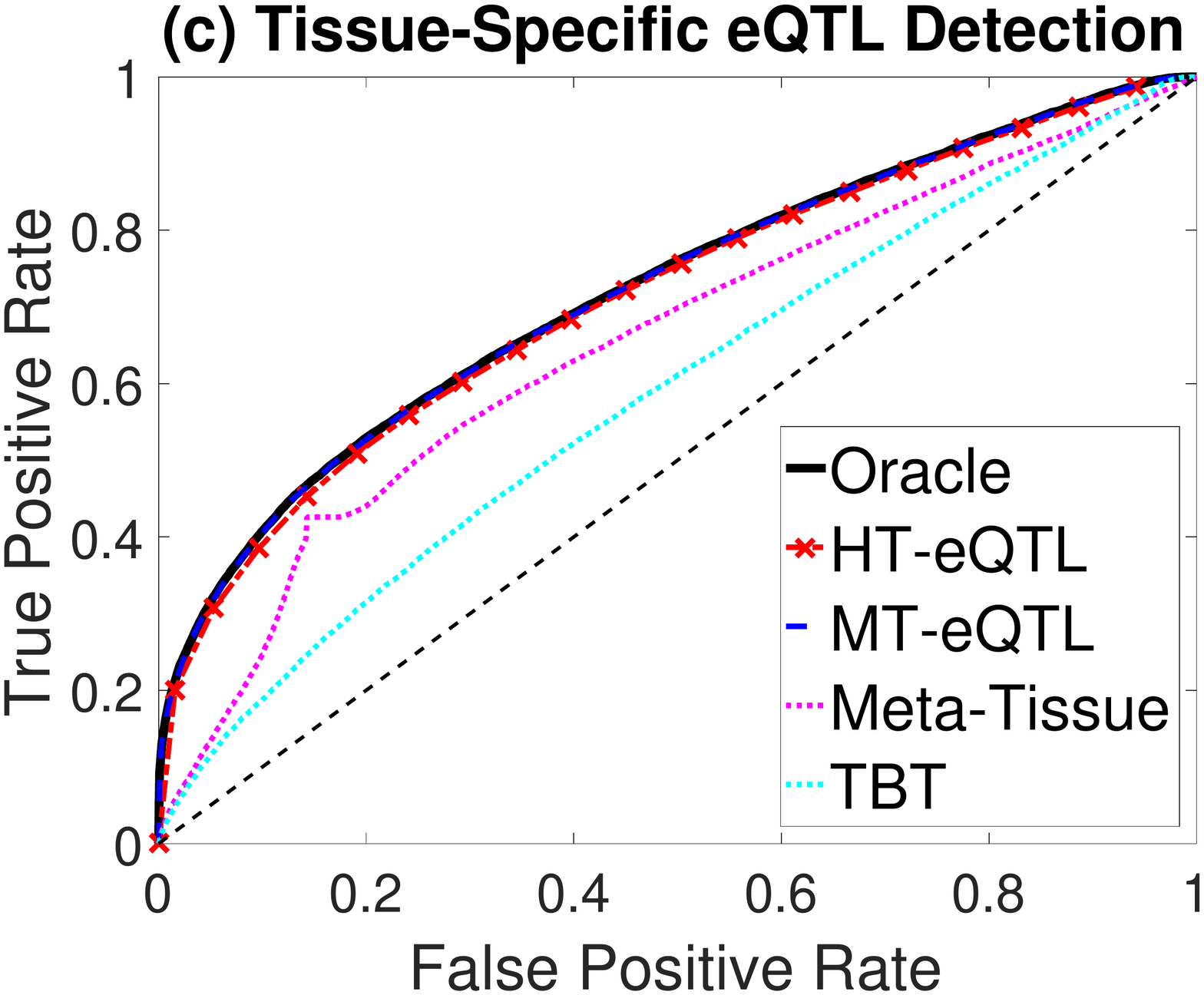}
\includegraphics[width=2.3in]{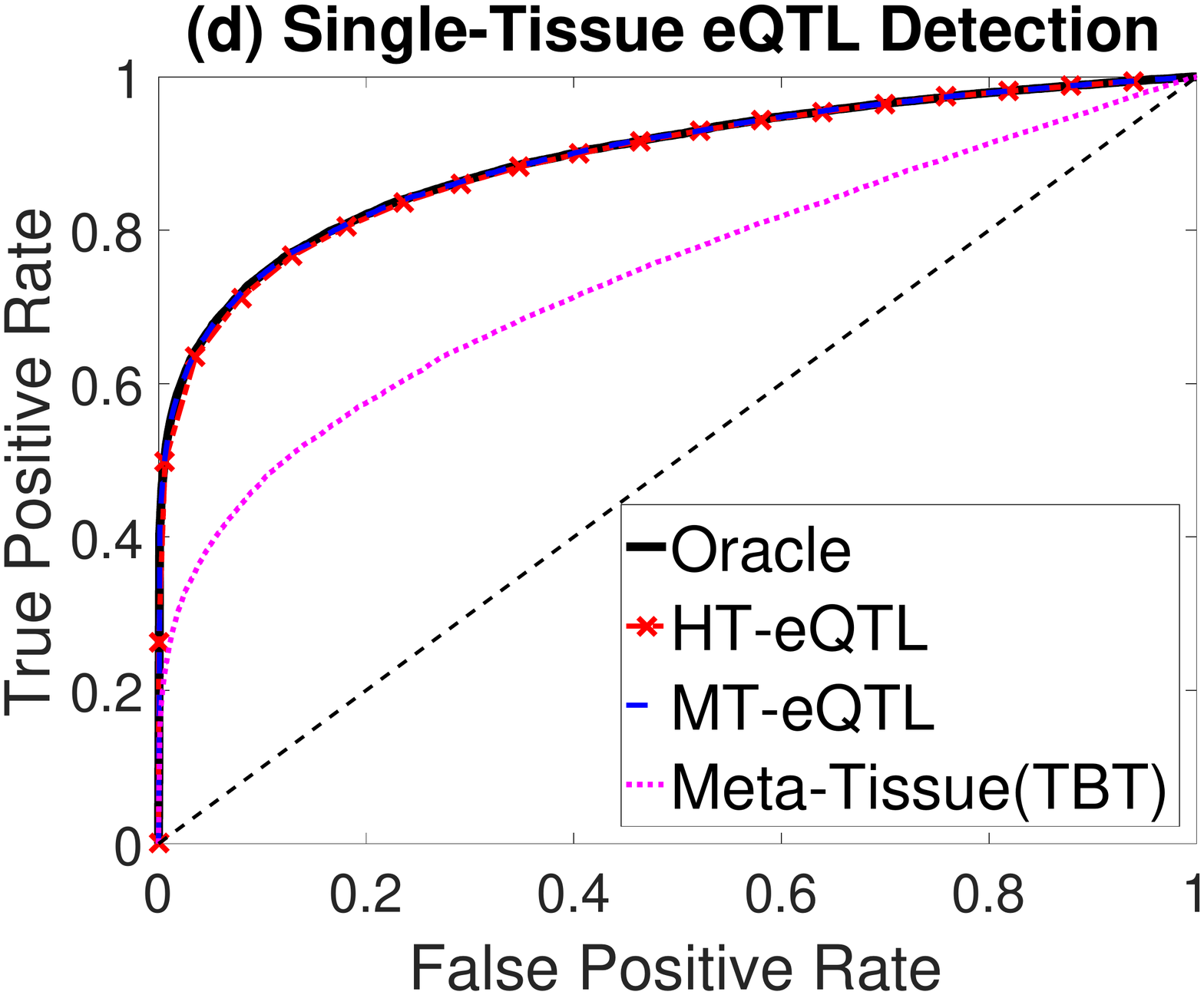}
\caption{The ROC curves of different methods for different eQTL detection problems in the simulation study. (a) Any eQTL detection; (b) Common eQTL detection; (c) Tissue-specific eQTL detection; (d) Single-tissue eQTL detection.}\label{fig:1}
\end{figure}

Now we compare the parameter estimation from MT-eQTL and HT-eQTL.
We particularly focus on the 9-tissue model.
The HT-eQTL parameters are obtained by fitting all 2-tissue MT-eQTL models and assembling the pairwise parameters. 
The MT-eQTL parameters are obtained directly by fitting the 9-tissue MT-eQTL model.
Regarding the estimation of the correlation matrix $\bDelta$, the quartiles of the entry-wise relative errors are (0.86\%, 2.42\%, 4.36\%) and (0.81\%, 2.00\%, 2.72\%) for HT-eQTL and MT-eQTL, respectively.
Regarding the estimation of the covariance matrix $\bSigma$, the quartiles of the entry-wise relative errors are (1.13\%, 2.41\%, 3.25\%) and (0.36\%, 0.68\%, 1.08\%) for HT-eQTL and MT-eQTL, respectively.
Namely, both methods estimated the covariance matrices very accurately.
For the probability mass vector $\bp$, we calculated the Kullback-Liebler divergence
of different estimates from the truth, defined as
$D_{KL}(\bp\|\widehat{\bp})=\sum_{i=1}^{2^K}p_i\log{(p_i/\widehat{p_i})}$.
The MT-eQTL estimate has a very small divergence of $0.025$ while the HT-eQTL estimate has a slightly larger divergence of $0.141$. 
Overall, the HT-eQTL estimates are slightly less accurate than the MT-eQTL estimates, which
is expected because the HT-eQTL method has fewer degrees of freedom than the MT-eQTL method.
When there are abundant data relative to the number of parameters, the more complicated
MT-eQTL model will result in more accurate estimation.
Nevertheless, we emphasize that the HT-eQTL estimates are sufficiently accurate for the eQTL detection purposes (see Figure \ref{fig:1}).

Next, we compare the eQTL detection power of different methods.
{\color{black} We particularly focus on the detection of four types of eQTLs: (a) eQTLs in at least one tissue (Any eQTL); (b) eQTLs in all tissues (Common eQTL); (c) eQTLs in at least one tissue but not all tissues (Tissue-Specific eQTL); (d) eQTLs in a single tissue (Single-Tissue eQTL).
In addition to the MT-eQTL and HT-eQTL methods, we also consider the Meta-Tissue and TBT approaches.
In order to detect Any eQTL, we exploit the random effects model in Meta-Tissue and a {\sf minP} procedure in TBT, where the minimum p value across tissues is used as the test statistics for each gene-SNP pair.
To detect Common eQTL, we use the fixed effects model in Meta-Tissue and a {\sf maxP} procedure in TBT, where the maximum p values across tissues are used.
To detect Tissue-Specific eQTL, we devise a {\sf diffP} procedure for TBT, where the test statistics for each gene-SNP pair is the difference between the maximum and the minimum p values across tissues.
A large value indicates the discrepancy between the two extreme p values is large, and thus
provides a strong evidence for the gene-SNP pair to be a tissue-specific eQTL.
Similarly, for Meta-Tissue, we exploit the difference of p values from the fixed effects model and the random effects model as the test statistics.
Finally, for Single-Tissue eQTL detection, Meta-Tissue reduces to the TBT method.
We just use the p values in the primary tissue and ignore those in other tissues.
For the MT-eQTL and HT-eQTL methods, we adapt the lfdr test statistics in \eqref{lfdr} to different testing problems accordingly.}

We evaluate the performance of different methods using the Receiver Operating Characteristic
(ROC) curves for different eQTL detection problems.
The results are shown in Figure \ref{fig:1}.
The oracle curves correspond to the lfdr approach based on the true model with the true parameters.
{\color{black}In all eQTL detection problems, the MT-eQTL and HT-eQTL methods have comparable performance, very similar to the oracle results.
While we expect the MT-eQTL to perform similarly to the oracle procedure, it is surprising that the HT-eQTL, only using information in tissue pairs, also provides comparable results to the oracle procedure.
Both MT-eQTL and HT-eQTL clearly outperform the Meta-Tissue and TBT approaches in all detection problems.}



To sum up, the HT-eQTL method achieves high parameter estimation accuracy and eQTL detection power at a low computational cost.
For a large number of tissues, it provides a preferable alternative to the MT-eQTL method.

\subsection{GTEx v6p Data}
The GTEx v6p data constitute the most recent freeze for official GTEx Consortium publications, and  can be accessed from the GTEx portal at \url{http://www.gtexportal.org/home/}.
{\color{black}We apply the HT-eQTL method to 20 tissues (selected by the GTEx Analysis Working Group), including 2 brain tissues, 2 adipose tissues, and a heterogeneous set of 16 other tissues.}
We consider all $70,724,981$ cis gene-SNP pairs where the SNP is within 1Mb of the TSS of the gene.

\begin{figure}[h!]
\centering
\includegraphics[width=2.5in]{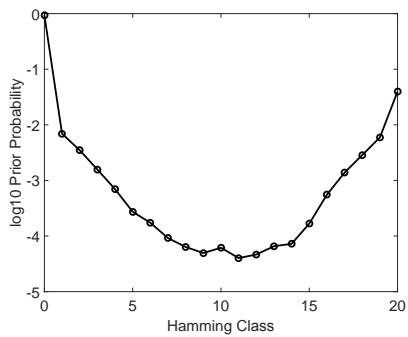}
\caption{The summary plot of the probability mass vector estimated from the HT-eQTL method on the GTEx v6p 20-tissue data. The prior probabilities are added up for configurations in the same Hamming class and then log-transformed. }\label{fig:4}
\end{figure}

To obtain model parameters using HT-eQTL, we first fit ${20 \choose 2}=190$ 2-tissue models, and then assemble all the pairwise parameters following the procedure in the method section.
The probability mass vector $\bp$ estimated from the Multi-Probit model is summarized in Figure \ref{fig:4}.
We particularly focus on 377 configurations with prior probabilities greater than $10^{-5}$.
{\color{black} The prior probabilities are added up for configurations in the same Hamming class,
providing a general characterization of the multi-tissue eQTL distribution.
The parabolic shape estimated from the data is concordant with previous results from the pilot study \cite{ardlie2015genotype}. }
The global null configuration (the binary $\0$ vector) has the largest probability
of $0.9359$, and the common eQTL configuration (the binary $\1$ vector) has the second largest probability
of $0.0396$.  Configurations in Hamming class 1 (eQTL in only one tissue) and 19 (eQTL in all but one tissues) have relatively large probabilities.  All other configurations have much lower probabilities.

\begin{figure}[h!]
\centering
\includegraphics[width=3in]{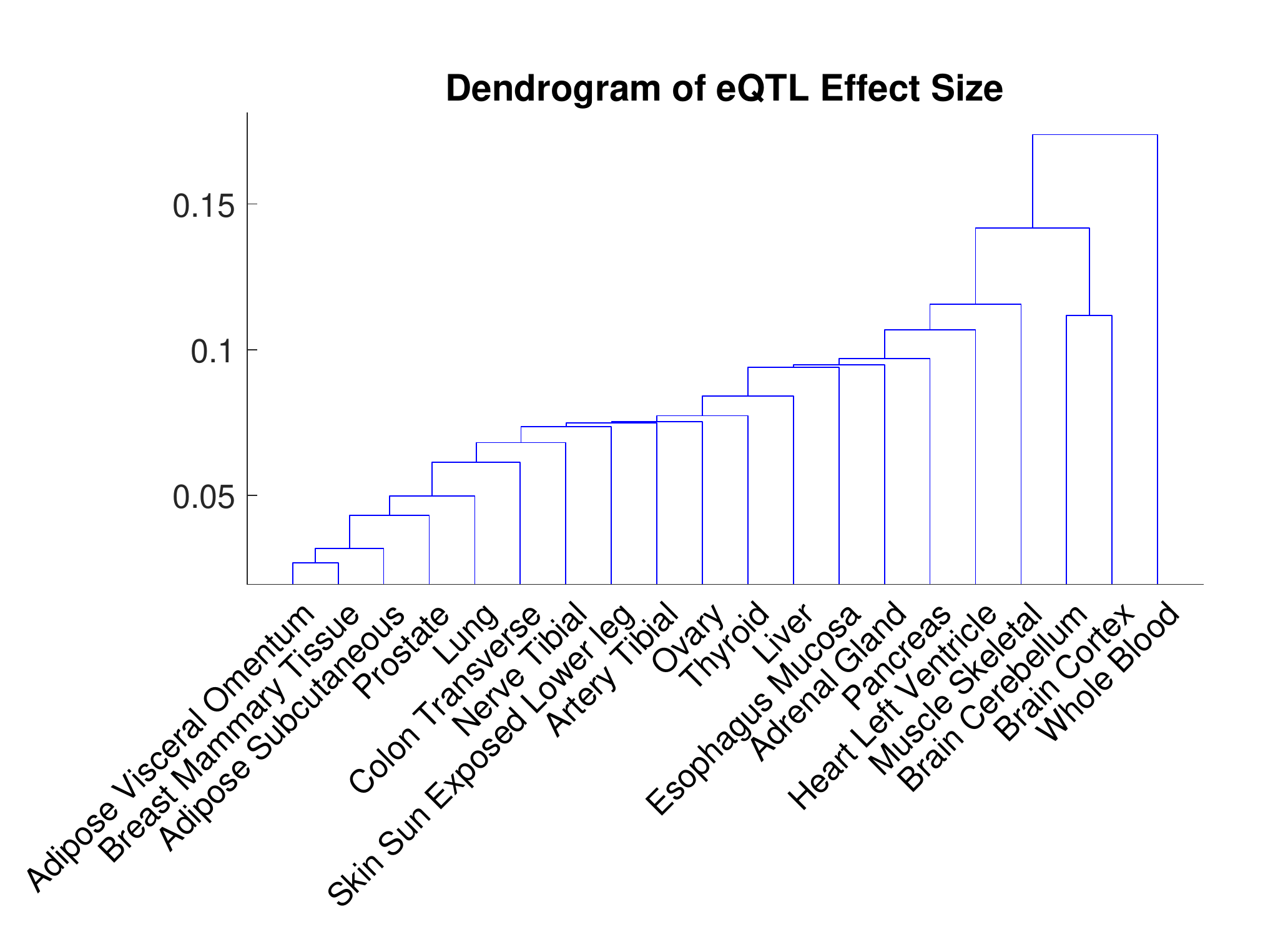}
\caption{The clustering result of 20 tissues in the GTEx v6p data analysis. The distance metric is the correlation of eQTL effect sizes between tissues, estimated from the HT-eQTL method.}\label{fig:5}
\end{figure}

Recall that $\bSigma$ captures the covariance of effect sizes in different tissues when eQTLs are present.
We treat the correlation matrix induced from $\bSigma$ as the distance metric between tissues, and use the single linkage to conduct hierarchical clustering for the 20 tissues.
The dendrogram is shown in Figure \ref{fig:5}.
We demonstrate that similar tissues, such as the two adipose tissues and the breast tissue, or the two brain tissues, are grouped together.
The whole blood is apparently different from all the other tissues.
These findings are concordant with those in the pilot analysis \cite{ardlie2015genotype}.

\begin{figure}[h!]
\centering
\includegraphics[width=2.5in]{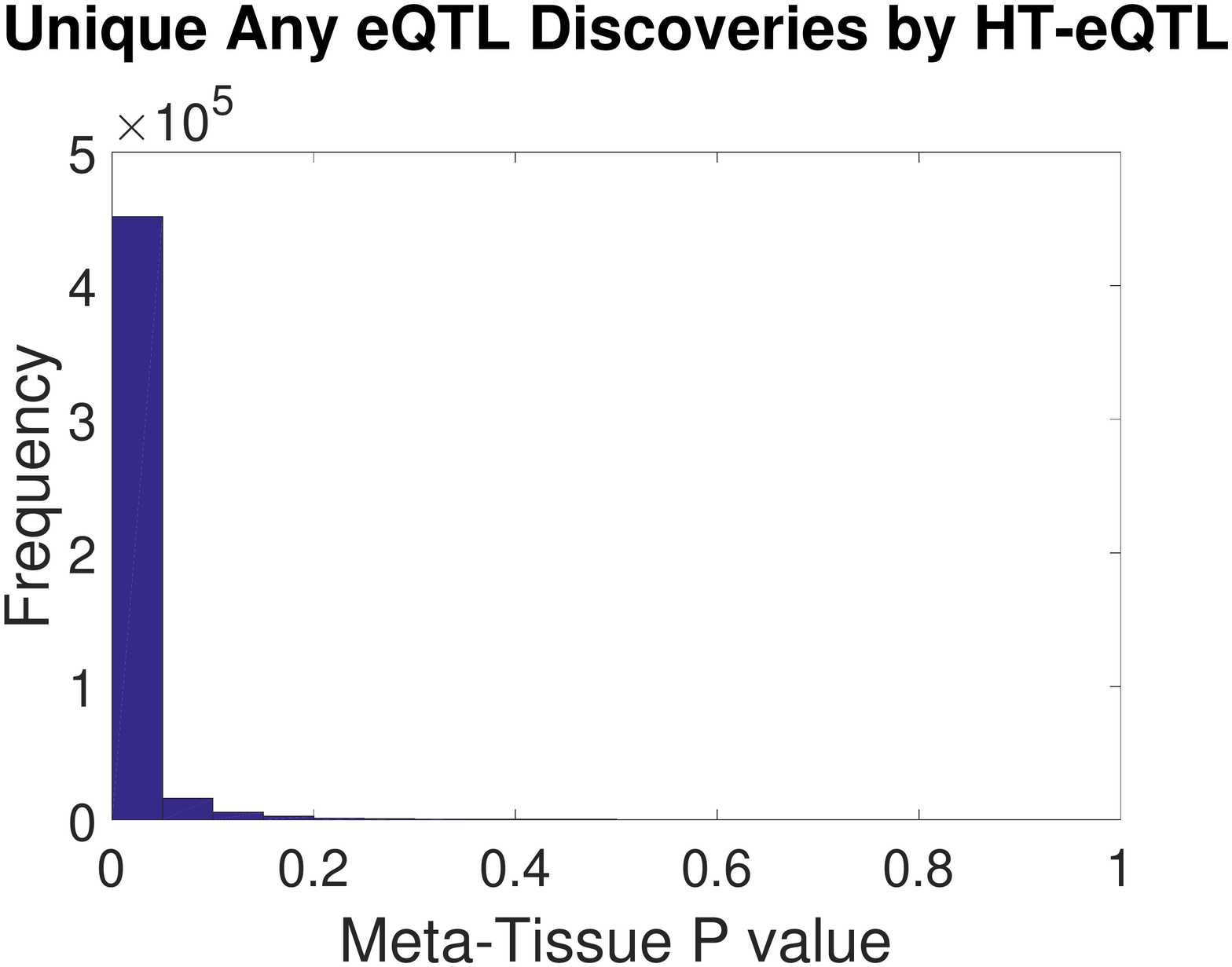}
\includegraphics[width=2.5in]{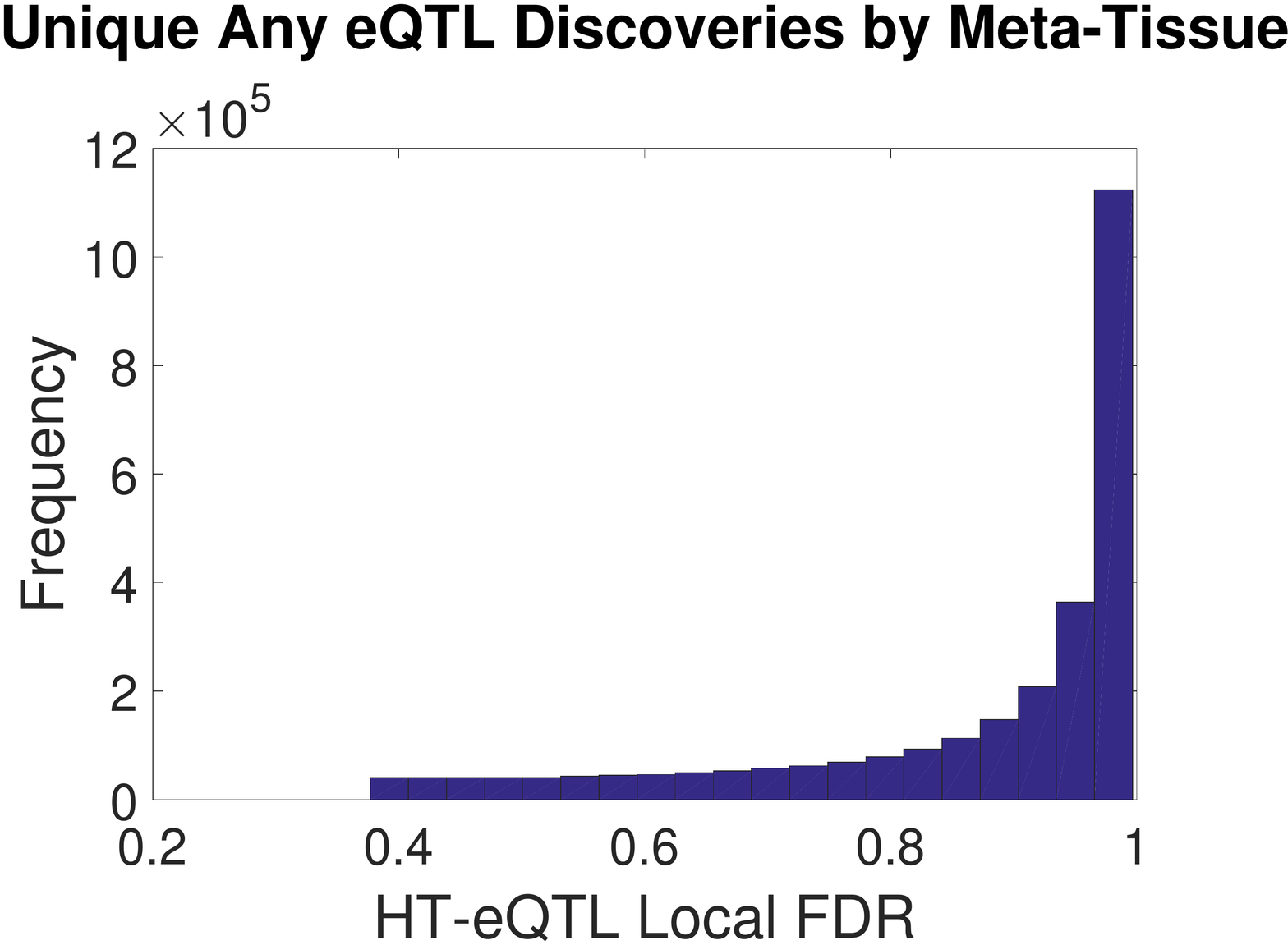}
\caption{Histograms of the Meta-Tissue p values for the unique Any eQTL discoveries made by HT-eQTL (left), and the HT-eQTL lfdr for the unique Any eQTL discoveries made by Meta-Tissue (right).
}\label{fig:hist}
\end{figure}

{\color{black}We also carry out testing of eQTL configurations (at a fixed the FDR level of $5\%$) for
the presence of an eQTL in any tissue, in all tissues, in at least one but not all tissues, and in each individual tissue.}
The number of discoveries are shown in Table \ref{tab:2}.
As a comparison, we also apply the Meta-Tissue method \cite{sul2013effectively} to the same data set.
{\color{black}In particular, we focus on the Any eQTL detection problem, using p values from the random effects model in Meta-Tissue.
We apply the Benjamini and Yekutieli approach \cite{benjamini2001control} to control the FDR at the level of 5\%.
As a result, we obtain over 6.36 million cis pairs from the Meta-Tissue method.
About 3.60 million of these pairs are shared with the HT-eQTL method.
We further investigate the unique discoveries of each method.
As shown in the left panel of Figure \ref{fig:hist}, the unique discoveries made by HT-eQTL have very small p values from the Meta-Tissue method, indicating those are likely to be ``near" discoveries for the Meta-Tissue method as well.
In the right panel of Figure \ref{fig:hist}, however, the excessive unique discoveries made by Meta-Tissue have highly enriched large lfdr values.
This suggests the unique Meta-Tissue discoveries may have inflated false discovery rate, potentially due to the inadequacy of the p-value-based FDR control method for highly dependent tests.
}

{\small
\begin{table}[h!]
\caption{The numbers of discoveries and the corresponding percentages of total cis pairs for different eQTL detection problems.  The FDR level is fixed at $5\%$ for all testing problems. \label{tab:2}}
\begin{center}
\begin{tabular}{ccc}
\hline
{\bf eQTL Configuration} & Number ($\times$1E6) & Percentage (\%) \\ \hline
eQTL in {\em Any} Tissue & 4.088 & 5.78\\
eQTL in {\em All} Tissues & 0.708 & 1.00 \\
{\em Tissue-Specific} eQTL & 0.239  & 0.34 \\ \hline
Adipose Subcutaneous & 3.640 & 5.15 \\ \hline
Adipose Visceral Omentum &3.536 & 5.00\\ \hline
Adrenal Gland & 3.302& 4.67\\ \hline
Artery Tibial & 3.671 & 5.19\\ \hline
Brain Cerebellum& 3.329 &4.71 \\ \hline
Brain Cortex & 3.120 & 4.41\\ \hline
Breast Mammary Tissue & 3.507 &4.96\\ \hline
Colon Transverse & 3.515 & 4.97\\ \hline
Esophagus Mucosa & 3.716 & 5.25 \\ \hline
Heart Left Ventricle &  3.433 & 4.85\\ \hline
Liver& 1.727 &2.44 \\ \hline
Lung & 3.576 & 5.06 \\ \hline
Muscle Skeletal & 3.581 &5.06\\ \hline
Nerve Tibial & 3.712 &5.25 \\ \hline
Ovary & 2.999 &4.24\\ \hline
Pancreas &3.479 &4.92\\ \hline
Prostate & 3.021 &4.27\\ \hline
Skin Sun Exposed Lower Leg & 3.717 &5.26\\ \hline
Thyroid & 3.758 &5.31\\ \hline
Whole Blood& 3.147 &4.45\\ \hline
\end{tabular}
\end{center}
\end{table}
}

\section{Discussion}
In this paper, we develop a new method, HT-eQTL, for joint analysis of eQTL in a large number of tissues.
The method builds upon the empirical Bayesian framework proposed in \cite{li2013empirical}, but is significantly improved in computation and inference to accommodate a large number of tissues.
The model fitting procedure only involves the estimation of all 2-tissue models, and the obtained pairwise parameters are then assembled to get the full model parameters.
The detection of eQTLs with different configurations is addressed by adaptively thresholding the corresponding local false discovery rates, which efficiently borrow strength across tissues and control the nominal FDR.
Finally, the numerical studies demonstrate the efficacy of the proposed method.
In the GTEx v6p data analysis, we apply HT-eQTL to 20 tissues.
The estimated prior probabilities of eQTL configurations show that most eQTLs are common across all tissues or present in a single tissue.
The estimated effect sizes provide additional insights into the tissue similarity and clustering.
We identify a large number of common and tissue-specific eQTLs.
A large proportion of the discoveries are replicated by the Meta-Tissue approach.
{\color{black} The additional unique discoveries made by our method are ``near" discoveries for the Meta-Tissue method, as illustrated by the highly skewed p-value distributions (see Figure \ref{fig:hist}).
It indicates that HT-eQTL is able to push the detection boundary in a favorable direction (i.e., more statistical power) while preserving error control.}

The proposed method relies on the marginal compatibility of the hierarchical Bayesian model \eqref{MT-model}.
In practice, if the joint distribution of the z-statistics deviates from  a multivariate Gaussian distribution, it may affect the model fitting.
One way to alleviate the problem is to transform the original z-statistics to make them jointly Gaussian.
A multivariate testing and transformation framework calls for more investigation.
{\color{black}Another limitation of HT-eQTL in its current form is that it is limited to 20-25 tissues.
Extensions beyond this number will require additional prior information about tissue groups, which may reduce the total number of configurations considered in the model.}

\section{Conclusions}

We present a scalable method for multi-tissue eQTL analysis.
The method can effectively borrow strength across tissues to improve the power of eQTL detection in a single tissue.
It also has superior power to detect eQTL of different configurations.
The model parameters capture important biological insights into tissue similarity and specificity.
In particular, from the GTEx analysis we observe that most cis eQTLs are present in either all tissues or a single tissue.
The eQTLs identified by the proposed method provide a valuable resource for subsequent analysis, and may facilitate the discovery of genetic regulatory pathways underlying complex diseases.

\section*{Acknowledgements}
The authors would like to thank members of the GTEx Analysis Working Group for helpful comments and discussions.

\section*{Funding}
 This work was supported by the National Institutes of Health [R01MH101819 and R01MH090936 to GL, ABN, FAW, R01HG009125 to ABN]; the National Science Foundation [DMS-1613072 to ABN]; and the National Institute of Environmental Health Sciences [P42ES005948 to FAW, P30ES025128 to DJ].

\end{document}